\documentclass[a4paper,11pt]{article}
\pdfoutput=1 

\usepackage{jheppub} 

\usepackage[T1]{fontenc}

\title{\boldmath A study of neutron stars in $D \ge 4$ dimensions}

\author[a,b, 1]{M.Bagchi,\note{Corresponding author.}}

\affiliation[a]{The Institute of Mathematical Sciences,
C. I. T. campus, Taramani, Chennai, 600113, India}
\affiliation[b]{Homi Bhabha National Institute, Anushakti Nagar, Mumbai 400094, India}

\emailAdd{manjari@imsc.res.in}

\abstract{The relativistic equations of hydrostatic equilibrium for a spherically symmetric star, or the Tolman-Oppenheimer-Volkoff equations are known in higher dimensions. In this paper, these equations have been expressed in terms of parameters of 4 dimensional spacetime and solved numerically for 4, 5, 6, and 7 dimensions using a standard equation of state for the neutron star matter derived for the 4 dimensional spacetime. It has been shown that with the increase of the dimensionality, the maximum value of the mass of the neutron star decreases and the stars become less compact. Thus, although the compactness limit decreases with increased dimensionality, neutron stars never violate this limit. Simultaneous measurements of the mass, radius, and gravitational redshift for a neutron star might enable us to conclude about the central density, equation of state and the dimensionality of the spacetime in and around the star.}

\begin{document} 
\maketitle
\flushbottom

\section{Introduction}
\label{sec:intro}

Many theories allow existence of higher dimensions that are non compact. Study of stability and structure of stars in such a framework is interesting and explored by different groups \cite{chav08, krama14, pc00}. All these studies have been performed for compact objects like white dwarfs, neutron stars, and black holes; all of which exhibit strong field gravity. These studies were mostly analytical and did not give numerical values of observable properties like the mass, radius, gravitational redshift, etc. for the stars. The reason for this is the fact that we do not know the value of the gravitational constant and properties of matter at higher dimensions. In the present work, I provide values of these observables for a few dimensions and discuss possible observational aspects.

\section{Metric, equations of hydrostatic equilibrium, and structure of static, isotropic, spherically symmetric stars in $D$ dimension}
\label{sec:formalism}

Spacetime  coordinates in $D$ dimension can be chosen as ($ct$, $x^i$, $r$, $\theta^j$) where $i= 1, 2, \ldots n_c$ represent the directions in the compact space, and $j= 1,2, \ldots m$ represent the directions in the non-compact transverse space, and $c$ is the speed of light in vacuum \cite{krama14} . Hence $D=2+n_c + m$ and the most general line element is:

\begin{equation}
\label{eq:line1}
ds^2 = e^{\nu} \, (c \, dt)^2 - \displaystyle\sum_{i=1}^{n_c}  e^{\mu_i} \,(dx^i)^2 - e^{\lambda} \, dr^2 - e^{\sigma} \, d\Omega_{m}^2 ~,
\end{equation} where $d\Omega_{m}^2 = d\theta_1^{\,2} + \sin^2 \theta_1 \, d\theta_2^{\,2} + \sin^2 \theta_1 \, \sin^2 \theta_2 \, d\theta_3^{\,2} + \ldots  \prod_{j=1}^{m-1} \sin^2 \theta_j \, d \theta_m^{2}$ is the line element on the $m$ dimensional unit sphere.

Ignoring the compact space, i.e., by taking $n_c =0$ one gets $D=2+m=3+n$ where we have introduced a new number $n=m-1=D-3$. For spherically symmetric metric in such a situation, eq. (\ref{eq:line1}) reduces to:

\begin{equation}
\label{eq:line2}
ds^2 = e^{\nu(r)} \, (c \, dt)^2 - e^{\lambda(r)} \, dr^2 - r^2 \, d\Omega_{m}^2 ~.
\end{equation}

Einstein's fields equations can be written as \cite{pc00}:

\begin{equation}
\label{eq:einstein}
R_{\alpha \beta} = \frac{8 \pi \widetilde{G}}{c^4} \left[ T_{\alpha \beta} -\frac{1}{n+1} g_{\alpha \beta} \, T \right] ~,
\end{equation} where the Greek indices ($\alpha$, $\beta$) go from 1 to $D$, $\widetilde{G}$ is the gravitational constant in $D$ dimension in the unit of $G ~{\rm length}^{D-4}$, and $G$ is the gravitational constant for $D=4$ dimensional spacetime. $T_{\alpha \beta}$ is the energy momentum tensor in the $D$ dimension.

For the isotropic distribution, the energy-momentum tensor becomes

\begin{equation}
\label{eq:energymomentum}
T_{\alpha}^{~ \beta} = diag(\widetilde{\rho c^2 }, -\widetilde{P}, -\widetilde{P}, \ldots, -\widetilde{P}) ~,
\end{equation} where $\widetilde{\rho} c^2$ is the energy density and $\widetilde{P}$ is the pressure in D dimension, both having the unit of mass-length$^{-D+3}$-time$^{-2}$.  $\widetilde{\rho}$ is the mass density in the unit of mass-length$^{-D+1}$.  The relation between $\widetilde{\rho c^2 }$ and $\widetilde{P}$ is the Equation of State (EoS) in $D$ dimension. The lack of numerical analysis for $D > 4$ is the fact that the values of $\widetilde{G}$, $\widetilde{\rho}\, c^2$, and $\widetilde{P}$ in various dimensions are not known. 

Solving eqns. (\ref{eq:einstein}) with the help of eqns (\ref{eq:energymomentum}), one gets \cite{pc00}:

\begin{subequations}\label{eq:fieldall}
\begin{align}
\label{eq:field1}
e^{-\lambda(r)} \left( \frac{\lambda^{\prime} }{r}  -\frac{n}{r^2} \right) + \frac{n}{r^2} & = \frac{16 \pi \, \widetilde{G} }{(n+1) c^4} \, \widetilde{\rho} c^2
 = \frac{16 \pi \, G }{(n+1) c^4} \, \rho c^2
\\
\label{eq:field2}
e^{-\lambda(r)} \left( \frac{\nu^{\prime} }{r}  + \frac{n}{r^2} \right) - \frac{n}{r^2} & = \frac{16 \pi \, \widetilde{G}}{(n+1) c^4} \, \widetilde{P} 
 = \frac{16 \pi \, G }{(n+1) c^4}  \, P
\\
\label{eq:field3}
e^{-\lambda(r)} \left( \frac{ \nu^{\prime \, \prime}}{2} +  \frac{{\nu^{\prime}}^2}{4}  -\frac{\nu^{\prime}  \lambda^{\prime} }{4} -\frac{  n \, \lambda^{\prime} + \nu^{\prime}}{2 \, r}  - \frac{n}{r^2} \right) + \frac{n}{r^2}  & = 0
\end{align} 
\end{subequations} where a prime symbol over a parameter means the first derivative with respect to $r$ and a double prime symbol means the second derivative with respect to $r$. In eqns. (\ref{eq:fieldall}), we have used $\widetilde{G} \, \widetilde{\rho} = G \, \rho$ and $\widetilde{G} \, \widetilde{P} = G \, P$ where $\rho$ is the mass density and $P$ is the pressure for $D=4$.  Integrating equations (\ref{eq:field1}), we get

\begin{equation}
\label{eq:metric1}
e^{-\lambda(r)} = 1 - \frac{2 G}{c^2}\, \cdot \, \frac{1}{r^n} \, m(r) ~,
\end{equation} where

\begin{equation}
\label{eq:massfunction}
m(r) = \frac{8 \pi}{n+1} \int_{0}^{r} \rho(r^{\prime}) \, {r^{\prime}}^{n+1} \, dr^{\prime}
\end{equation} is called the mass-function. Note that the dimension of $m(r)$ is not the dimension of mass, it is mass-length$^{n-1}$, and the density is kept inside the integration as inside a star the density is a function of the radial coordinate.

The conservation of energy gives:

\begin{equation}
\label{eq:energycons}
\begin{split}
 \frac{d\widetilde{P}(r)}{dr} & = -\frac{1}{2} \frac{d \nu}{dr} \left( \widetilde{\rho} c^2 + \widetilde{P} \right) ~~~ {\rm or}
\\ 
\frac{dP(r)}{dr} & = -\frac{1}{2} \frac{d \nu}{dr} \left( \rho c^2 + P \right)
\end{split}
\end{equation}

Using eqns. (\ref{eq:field2}), (\ref{eq:metric1}), and (\ref{eq:energycons}), we get

\begin{equation}
\label{eq:dpdr}
\frac{dP(r)}{dr} = \frac{- \frac{G}{c^4} \left[ \rho(r) c^2 + P(r) \right] \, \left[ n(n+1) \, m(r) \, c^2 + 8 \pi \, P(r) \, r^{n+2} \right] }{(n+1) \, r^{n+1} \, \left[ 1 - \frac{2 G}{c^2} \cdot \frac{m(r)}{r^n} \right]} ~.
\end{equation} 

Eqns. (\ref{eq:dpdr}) and (\ref{eq:massfunction}) are the equations of hydrostatic equilibrium in $D$ dimension. For $D=4$, i.e., $n=1$, these equations reduces to the well known Tolman-Oppenheimer-Volkoff equations \cite{tol39, op39}.

To get the structure of a star, one first needs to choose a high density $\rho_c$ at the centre of the star ($r=0$) and corresponding pressure $P_c$ from a chosen EoS, and calculate $\Delta \, m = m(r=\Delta \, r)$ by assuming that the density is constant over a small $\Delta \, r$. Using these initial values of the pressure and mass-function one integrates eqns. (\ref{eq:dpdr}) and (\ref{eq:massfunction}) until the value of the pressure becomes zero. The value of $r$ at which one gets zero pressure is the radius $R$ of the star, and $m(r=R) = \mathcal{M}$ is the total ``mass'' of the star. $\mathcal{M}$ also has the dimension of $M \, L^{n-1}$, and we can define the physical mass of the star as $M=\mathcal{M}/R^{n-1}$ which has the dimension of mass. The line element interior and exterior of such a spherically symmetric static star can be written as \cite{pc00}:

\begin{equation}
\label{eq:starmetric}
\begin{split}
ds^2 &= \left( 1 - \frac{2G}{c^2} \, \frac{m(r)}{r^n} \right) \, (c \, dt)^2 - \left( 1 - \frac{2G}{c^2} \, \frac{m(r)}{r^n} \right)^{-1} \, dr^2 - r^2 \, d\Omega_{m}^2
\qquad
r < R 
\\
 &= \left( 1 - \frac{2G}{c^2} \, \frac{\mathcal{M}}{R^n} \right) \, (c \, dt)^2 - \left( 1 - \frac{2G}{c^2} \, \frac{\mathcal{M}}{R^n} \right)^{-1} \, dr^2 - r^2 \, d\Omega_{m}^2 \qquad
r \geq R 
\\
 &=  \left( 1 - \frac{2G}{c^2} \, \frac{M}{R} \right) \, (c \, dt)^2 - \left( 1 - \frac{2G}{c^2} \, \frac{M}{R} \right)^{-1} \, dr^2 - r^2 \, d\Omega_{m}^2
\qquad
r \geq R ~.
\end{split}
\end{equation}

From eqn (\ref{eq:starmetric}), it is obvious that the surface of the star would show a singularity if $1 \geq \frac{2G}{c^2} \, \frac{M}{R} $ or $ R \geq \frac{2G}{c^2} \, M$. This is independent of $n$ and same as the Schwarzschild limit for the case of $D=4$. Similarly, the gravitational redshift parameter can be written as:

\begin{equation}
\label{eq:redshift}
z+1 = \frac{\lambda_{\rm observed}}{\lambda_{\rm emitted}} = \left( 1 - \frac{2 G}{c^2} \cdot \frac{M}{R} \right)^{-1/2} ~,
\end{equation} which is also independent of $n$ and similar to the case of $D=4$.

The compactness limit for a spherically symmetric static star is known as \cite{pc00}:

\begin{equation}
\label{eq:buchdahl}
\frac{G}{c^2} \cdot \frac{M}{R} \leq \frac{2 \, (n+1)}{(n+2)^2} ~,
\end{equation} which reduces to the standard `Buchdahl limit' \cite{buch59} $\frac{G}{c^2} \cdot \frac{M}{R} \leq \frac{4}{9}$ for 4-dimensional spacetime ($n=1$). It is possible to numerically solve eqns. (\ref{eq:massfunction}), (\ref{eq:dpdr}), (\ref{eq:redshift}), and (\ref{eq:buchdahl}), as all of these now contain parameters from $D=4$, e.g., $G$, $\rho$, and $P$, which are known. In particular, we can choose a known EoS derived for $D=4$ to get values of $\rho$ and $P$. We report results of such numerical solutions in the next section.

\section{Static, isotropic, spherically symmetric neutron stars with an APR-like equations of state in $D$ dimension}
\label{sec:results} 

There are many EsoS for dense matter, out of which, many have been ruled out by the gravitational wave event GW170817 detected by the LIGO-Virgo collaboration \cite{aa18}. The assumption was that the event was caused by mergers of two neutron stars in $D=4$.  We choose one of the few EsoS allowed by the GW170817 event, i.e, the Akmal-Pandharipande-Ravenhall (APR) EoS \cite{apr98}.

We compute $M$ and $R$ for various central densities of stars for $n=1,2,3,4$ by numerically solving eqns. (\ref{eq:dpdr}) and (\ref{eq:massfunction}). The resulting $M-R$ relations are shown with solid lines in fig. \ref{fig:massradius} where the radius in kilometer is plotted along the abscissa and the mass in the unit of solar mass ($M_{\odot} = 1.98847 \times 10^{30}$ kg) is plotted along the ordinate. Each point on a $M-R$ curve represents a particular star and different points are obtained with different choice of the central density $\rho_c$. For any fixed $n$, a small value of $\rho_c$ gives a small mass and large radius. As the value of $\rho_c$ increases, the mass of the star also increases until it reaches a maximum value, this corresponds to $\rho_{c, max}$. No stable stellar configuration is possible for $\rho_c \geq \rho_{c, max}$.

We find that the $M-R$ curves always stay below the compactness limit given by eqn (\ref{eq:buchdahl}). This limit for each $n$ has been shown in the same figure with dashed lines. Moreover, lines for $z=$ 0.05, 0.1, 0.14, 0.2, 0.3, and 0.7 have been shown with dots, and the singular region has been shown as the hatched region in the top-left corner of the plot.

\begin{figure}
\centering
\includegraphics[width=0.7\textwidth, angle=-90]{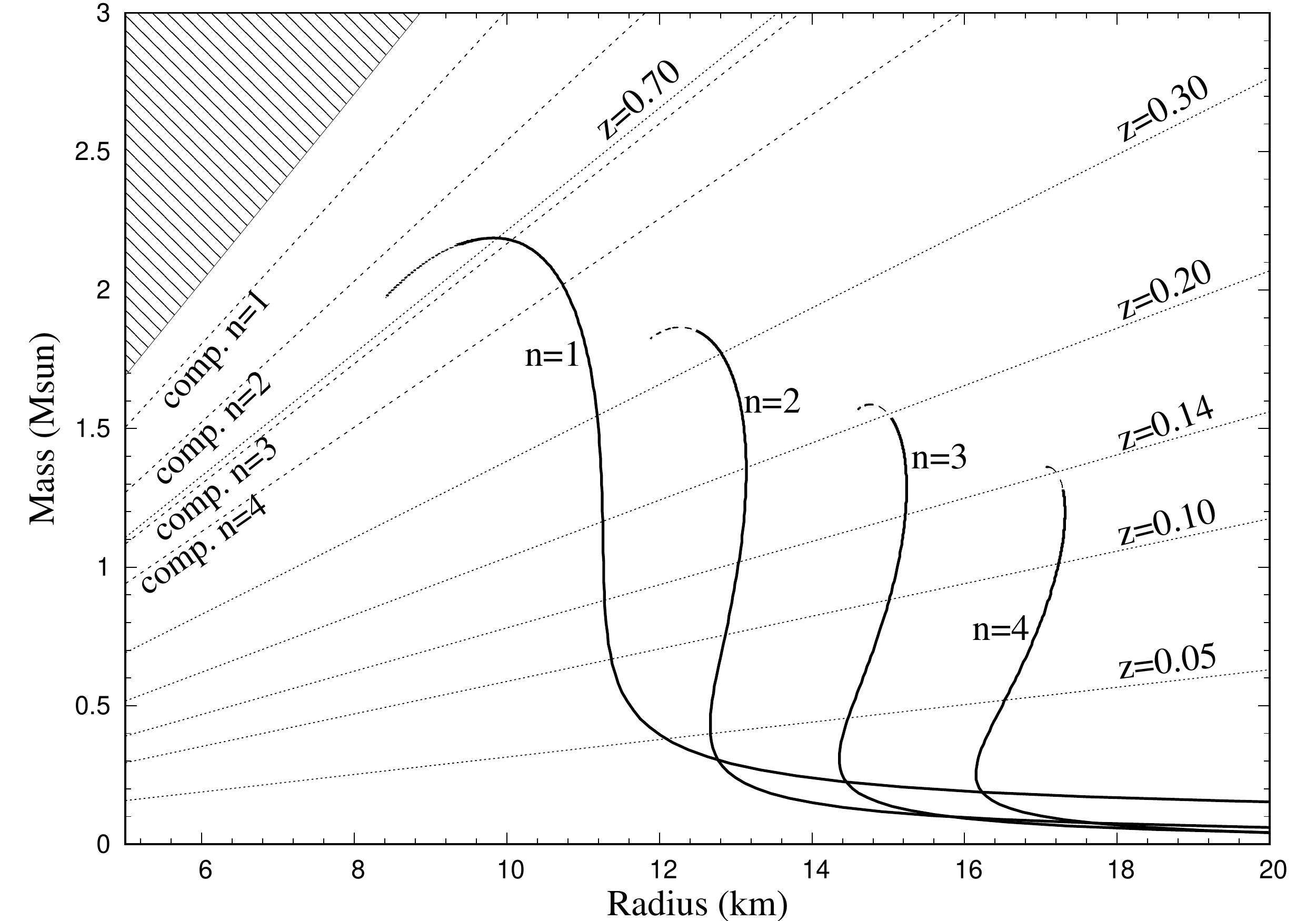}
\caption{Mass-radius plots (solid lines) for APR EoS obtained for $n=1, \,2, \,3, {\rm and} \, 4$ where $n=D-3$. The hatched region in the top-left corner is the region of singularity. The limit of compactness are shown for each $n$ with dashed lines. The lines for gravitational redshift parameter $z=$ 0.05, 0.1, 0.14, 0.2, 0.3, and 0.7 are shown with dotted lines.}
\label{fig:massradius}
\end{figure}

\begin{figure}
\centering
\includegraphics[width=0.7\textwidth, angle=-90]{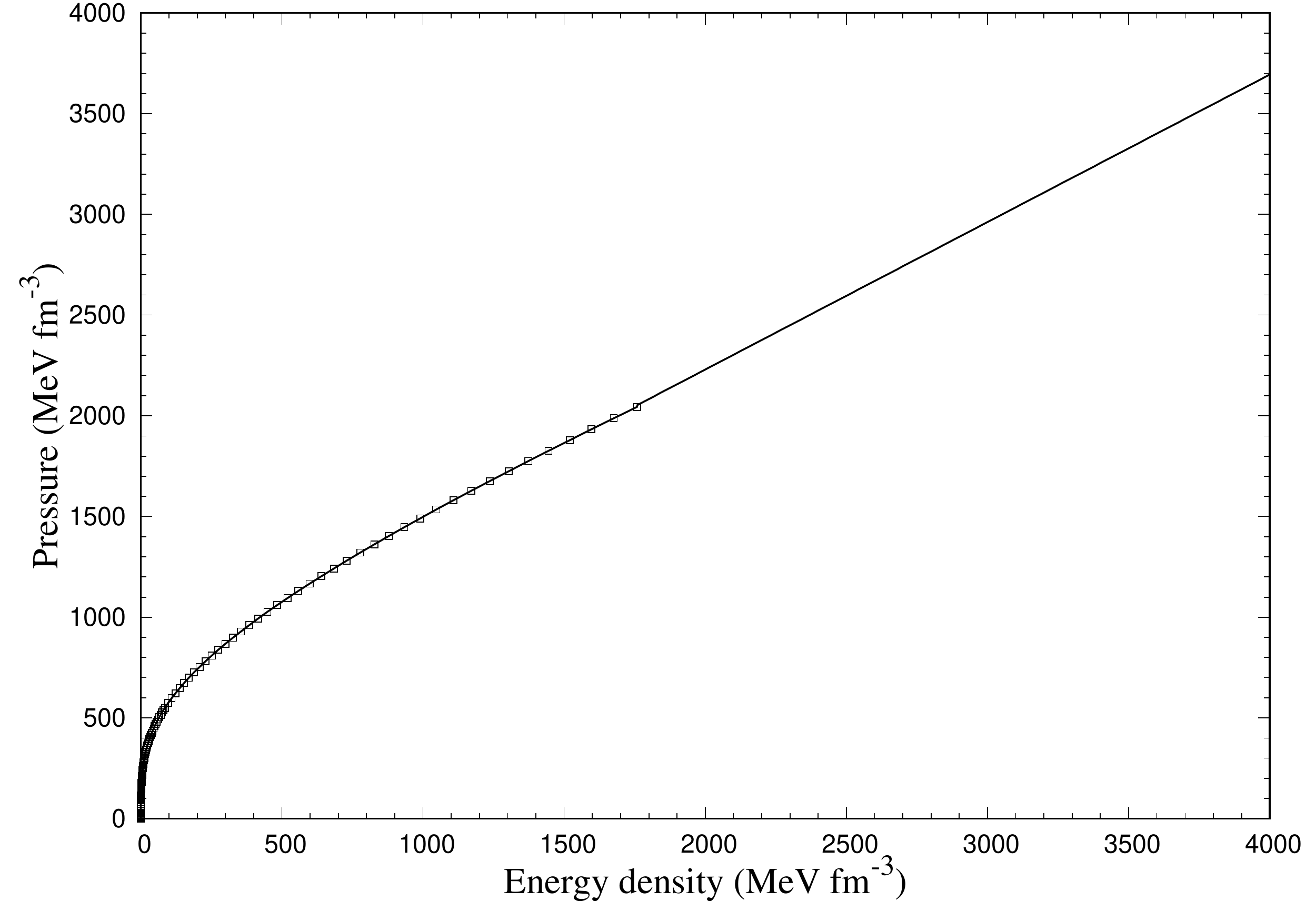}
\caption{APR EoS. The part marked with small squares is the EoS available in the literature and the straight line is our fit.}
\label{fig:massradius}
\end{figure}

Our most interesting findings are: (i) for each $n$, we obtain a maximum mass, (ii) the value of the maximum mass decreases with increasing $n$, (iii) the central density is higher for the maximum mass at higher dimension as shown in Table \ref{tab:maximummass}, (iv) the stars are less compact for larger $n$, e.g., for $M=1.03 \, {\rm M_{\odot}}$, we get $R=$ 11.27, 13.05, 15.14, and 17.25 km, and central densities 0.453$\times 10^3$, 0.663$\times 10^3$, 0.873$\times 10^3$, and 1.173$\times 10^3$  ${\rm~ MeV ~  fm^{-3}}$ for $n=$ 1, 2, 3, and 4 respectively.

Also, note that the APR EoS in the literature is available upto the density 2.0439 $\times 10^3 {\rm ~ MeV ~  fm^{-3}}$, which does not give the maximum mass for $n>2$. So, we have extrapolated the EoS by fitting the high density part with a straight line $P=0.732112 \times (\rho \, c^2) + 765.528$. Figure 2 shows the EoS, both the original and the fit.

\begin{table}[tbp]
\centering
\begin{tabular}{|l |c |c |c |c|}
\hline
parameter & $n=1$ & $n=2$ & $n=3$ & $n=4$ \\
\hline 
maximum mass (${\rm M_{\odot}}$) & 2.19 & 1.84 & 1.57 & 1.35 \\
 &  &  &  & \\
corresponding radius (${\rm km}$) & 9.83 & 12.52 & 14.93 & 17.17 \\
 &  &  &  & \\
corresponding central density ($10^{3} \, {\rm MeV~fm^{-3}}$) & 1.553 & 1.963  & 2.573  &  3.783 \\
\hline
\end{tabular}
\caption{\label{tab:maximummass} Maximum mass and corresponding radius and central density for $n=$ 1, 2, 3, and 4 where $n=D-3$.}
\end{table}

\section{Discussion}
\label{sec:disc}

Note that, there are some neutron stars with measured masses larger than 2 ${\rm M_{\odot}}$, e.g., PSR J0740+6620 with $2.14^{+0.20}_{-0.18}~ {\rm M_{\odot}}$ \cite{cfr20} and PSR J0348+0432 with $2.01 \pm 0.04 ~ {\rm M_{\odot}}$ \cite{afw13}. This implies that at least for these two neutron stars, either we can rule out the possibility of $n \geq 2$ as the maximum mass possible for $n=2$ is $1.84~{ \rm M_{\odot}}$ and lower for higher $n$ for APR EoS, or the EoS is much stiffer which would result a higher value of the maximum mass for each value of $n$. However, it still remains an open question whether low mass neutron stars have just low central density, or they belong to higher dimensions. In principle, if one can measure $M$, $R$, $z$ - all three at the same time it would be possible to constrain both the EoS, central density and the dimension of the spacetime inside that object. This will be an extremely challenging task for observational astronomers, but not impossible by combining timing and spectral analysis of binary pulsars. Where timing analysis of binary pulsars would result in measurements of the mass of the star while the spectral analysis would give the radius (using the value of the mass obtained from timing analysis) and gravitational redshift (if any known spectral line is  detected).

\acknowledgments

The author thanks S. Kalyana Rama for many illuminating discussions.

\end{document}